\RequirePackage{amsthm}
\documentclass[sn-nature,iicol]{sn-jnl}% Double column layout

\usepackage{graphicx}%
\usepackage{multirow}%
\usepackage{amsmath,amssymb,amsfonts}%
\usepackage{mathrsfs}%
\usepackage[title]{appendix}%
\usepackage{xcolor}%
\usepackage{textcomp}%
\usepackage{manyfoot}%
\usepackage{booktabs}%
\usepackage{algorithm}%
\usepackage{algorithmicx}%
\usepackage{algpseudocode}%
\usepackage{listings}%

\theoremstyle{thmstyleone}%
\theoremstyle{thmstyletwo}%

\theoremstyle{thmstylethree}%

\begin{document}

\title{Universal proximity time index}
\title{Proximity-based cities enhance inclusion}
\title{A universal number of 15 minute cities}
\title{Investigating Accessibility in Global Cities: The Feasibility of the 15-Minute City Concept}
\title{The Viability of 15-Minute Cities: A Study on Urban Accessibility and Population Density}
\title{The Promise and Challenges of Global 15-Minute Cities}
\title{Walking Towards Equality: The Global Quest for 15-Minute Cities}
\title{Walking towards equal accessibility in the global quest for 15-minute cities}
\title{Rethinking local accessibility equality towards the 15-minute city}
\title{Equalising accessible opportunities in cities: the challenges towards 15-minute cities}
\title{Is mine a 15-minute city, or can it become one?}
\title{Some cities cannot become 15-minute cities}
\title{Which cities are 15-minute, and which ones cannot become one}
\title{A 15-minute city has to be compact}
\title{The 15-minute city and equal opportunities: a dream or a possibility?}
\title{It is time to develop cities on the inside and not grow them on the outside}
\title{Equal 15-minute cities: utopia or reality?}
\title{Towards the 15-minute city: designing equal accessibility}
\title{Towards inclusive proximity-based cities}
\title{Inclusiveness in proximity-based cities}
\title{A universal framework for inclusive 15-minute cities}

\author*[1,2]{\fnm{Matteo} \sur{Bruno}}\email{matteo.bruno@sony.com}
\author[1,2,3]{\fnm{Hygor Piaget} \sur{Monteiro Melo}}
\author[1,2,4]{\fnm{Bruno} \sur{Campanelli}}
\author[1,2,4,5]{\fnm{Vittorio} \sur{Loreto}}

\small{
\affil[1]{\orgname{Sony Computer Science Laboratories - Rome}, Joint Initiative CREF-SONY, Centro Ricerche Enrico Fermi, \orgaddress{\street{Via Panisperna 89/A}, \postcode{00184}, \city{Rome}, \country{Italy}}}
\affil[2]{\orgname{Centro Ricerche Enrico Fermi (CREF)}, \orgaddress{\street{Via Panisperna 89/A}, \postcode{00184}, \city{Rome}, \country{Italy}}}
\affil[3]{\orgname{Instituto Federal de Educação, Ciência e Tecnologia do Ceará}, \orgaddress{\street{Avenida Des. Armando de Sales Louzada}, \city{Acaraú}, \state{Ceará}, \country{Brazil}}}
\affil[4]{\orgname{Sapienza Univ. of Rome}, \orgdiv{Physics Dept}, \orgaddress{\street{Piazzale A. Moro, 2}, \postcode{00185}, \city{Rome}, \country{Italy}}}
\affil[5]{\orgname{Complexity Science Hub}, \orgaddress{Josefst\"{a}dter Strasse 39}, \postcode{A 1080} \city{Vienna}, \country{Austria}}
}

% \affil[5]{\orgname{Instituto de Fısica Interdisciplinar y Sistemas Complejos IFISC (CSIC-UIB)}, \postcode{07122} \city{Palma de Mallorca}, \country{Spain}

\abstract{Proximity-based cities have attracted much attention in recent years. The 15-minute city, in particular, heralded a new vision for cities where essential services must be easily accessible. Despite its undoubted merit in stimulating discussion on new organisations of cities, the 15-minute city cannot be applicable everywhere, and its very definition raises a few concerns. Here, we tackle the feasibility and practicability of the '15-minute city' model in many cities worldwide. We provide a worldwide quantification of how close cities are to the ideal of the 15-minute city. To this end, we measure the accessibility times to resources and services, and we reveal strong heterogeneity of accessibility within and across cities, with a significant role played by local population densities. We provide an online platform (\href{whatif.sonycsl.it/15mincity}{whatif.sonycsl.it/15mincity}) to access and visualise accessibility scores for virtually all cities worldwide. The heterogeneity of accessibility within cities is one of the sources of inequality. We thus simulate how much a better redistribution of resources and services could heal inequity by keeping the same resources and services or by allowing for virtually infinite resources. We highlight pronounced discrepancies among cities in the minimum number of additional services needed to comply with the 15-minute city concept. We conclude that the proximity-based paradigm must be generalised to work on a wide range of local population densities. Finally, socio-economic and cultural factors should be included to shift from time-based to value-based cities.}


\maketitle

% \linenumbers

\section{Introduction}

In the last few years, a strong push for redesigning our urban landscapes emerged to set the stage for a future of more sustainable and connected urban living. This novel movement took off from the need for local accessibility to services and reducing carbon emissions from private transportation. 

The recently trending concept of the '15-minute city' contends that cities can function more effectively, equitably, and environmentally if essential services and key amenities, are within 15 minutes by non-polluting forms of transport, such as walking or cycling~\cite{moreno2021introducing}. It provides an alternative to car-dependent urban designs and generally deleterious long-distance commuting. By focusing on proximity, this approach can address pressing issues like air pollution, traffic congestion, and social disparities~\cite{allam2022theoretical}. The 15-minute city concept has also significantly risen in popularity during and after the COVID-19 pandemic, as local communities regained importance and people, even reluctantly, rediscovered the possibility of enjoying their neighbourhood~\cite{basbas2023citymodel}, calling for new urban solutions to enhance walkability~\cite{rhoads2021sustainable}.

The notion of proximity-based cities has been around for a while, due to decades of increasing interest in compact cities~\cite{perry_neighborhood_1929,pozoukidou2022urban}. The long-standing dualism between suburban cities and compact urban planning has been at the centre of public debates recently due to the need to create sustainable environments and reduce CO2 produced by transport emissions, which recent studies confirm to be lower in dense cities~\cite{ribeiro2019effects}. Therefore, a new focus on eco-urbanism solutions has arisen, with the challenge of building more compact cities and improving accessibility~\cite{haaland2015challenges}.

Transforming a city into a 15-minute one is a challenging task. It calls for significant shifts in urban design, transportation strategies, and land use policies~\cite{khavarian202315}. Despite its increasing appeal, the 15-minuteness raises essential concerns. For once, a perfectly 15-minute city could be highly unequal regarding the quality of the services provided or might lead to a reduction of green areas~\cite{burton2000compact}. Two areas of the same city could have perfect 15-minuteness, but one could have top-tier and the other inferior services, forming a fertile ground for inequality and segregation. In addition, as we shall see in this paper, the local population density plays a significant role in determining the best scenarios for future cities. The practicality of the 15-minute city still needs to be determined, as well as its consequential impact on the resident's quality of life. Some studies have begun to quantitatively measure the impact and influence of such an urban structure~\cite{lima2023quest}. Moreover, many cities have begun journeys to improve accessibility and create compact districts ~\cite{pozoukidou202115}, but there is no unique recipe for such changes.

Measuring local accessibility to services becomes a cornerstone in this exploration. The quest for measuring accessibility began long ago, and various measures have been employed~\cite{levinson2020towards,levinson2020transport,cui2020primal}. All these studies point out that cities are not homogeneous bodies: it is crucial to acknowledge that not all parts of a city are equal, leading to varying degrees of accessibility~\cite{weiss2018global,biazzo2019general,liu2021studying,kaufmann2022scaling,bittencourt2023evaluating}; thus, previous studies point towards inequality of access even in 15-minute cities~\cite{yang2023visualizing,vale2023accessibility}. Therefore, urban studies have focused on population landscapes, i.e., the distributions of local population density and how to improve accessibility for the largest possible fraction of the population. One popular approach is to place amenities to meet the needs of the population optimally; this complex problem can be formulated in various ways, and some algorithms have been proposed~\cite{gastner2006optimal,xu2020deconstructing,lima2022grammar}, even considering mobility~\cite{fan2022equality}. Here, we aim to merge these theoretical accessibility designs with the quest for the 15-minute city to understand how optimally placed amenities can improve local accessibility in urban contexts.

As a first contribution of the paper, we measure accessibility in worldwide cities, probing the distribution of services and pinpointing regions where disparities are the most conspicuous. Further, we scrutinise the practicality of the '15-minute city' concept by imagining a more equal distribution of services, considering the spatial distribution of population densities.
We propose a novel algorithmic approach to create scenarios wherein city services are redistributed based on population distributions. By simulating a hypothetical equal relocation of services, we can forecast potential enhancements in urban accessibility. Finally, we employ our algorithm to fine-tune the number of services to reach an equal '15-minute city' condition. Measuring the quantity of per capita services needed for this goal, we demonstrate that the number of services required fluctuates remarkably and depends on the population density and how the population is distributed in the territory. We contend that our analysis can offer insights that can serve as catalysts for urban planners and decision-makers, fostering the creation of cities that are more equitable, accessible, and sustainable.

\section{Results}

\subsection{Assessing current 15-minuteness}

\begin{figure*}[htbp]
\includegraphics[width=\textwidth]{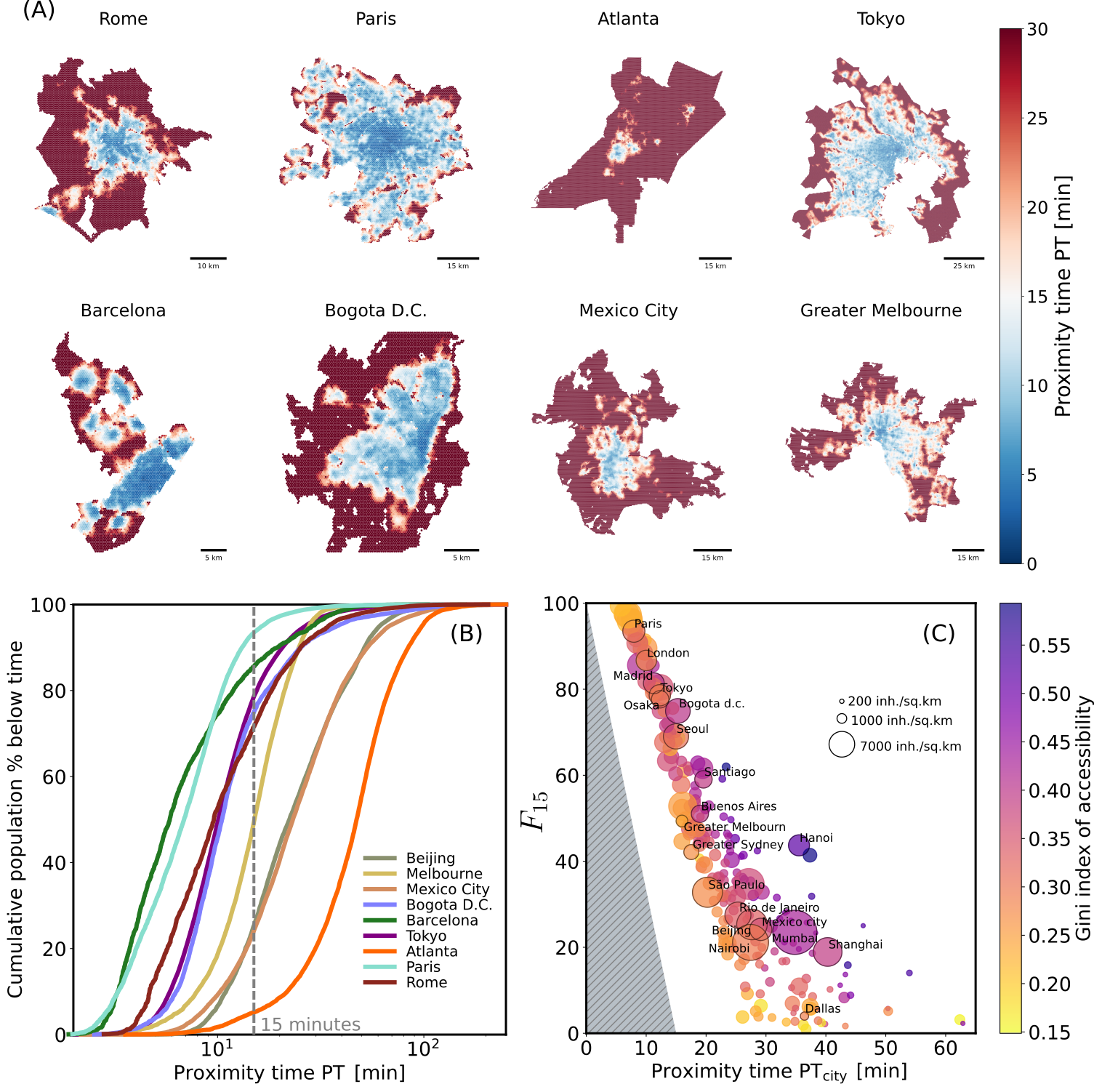}
\caption{\textbf{Local accessibility measured in several different cities.} {\bf Panel A} shows the computed maps of local accessibility scores for a subset of the studied cities. Each hexagon in each city is coloured according to the Proximity Time, i.e., the time to access the services in the basket of the 15-minute city. The colour code is such that blueish (reddish) colours correspond to areas whose accessibility time of the services is below (above) 15 minutes. Remarkable differences are observed among different cities. For instance, a small fraction of the city area in Atlanta has PTs that are less than 15 minutes. Conversely, Paris has a significant fraction of the city area below 15 minutes. {\bf Panel B} shows the cumulative distribution of PT scores among the population in selected cities, i.e., the population fraction whose accessibility is below a given PT. The curves for different cities display similar S-shaped behaviours, though their growth occurs at very different positions. We highlight, in particular, the curve intercept for a PT of 15 minutes, revealing the fraction of the population living in a 15-minute condition. The unequal accessibility across cities is captured by the Gini index depicted in panel C. {\bf Panel C} shows a scatter plot of different accessibility scores of cities in our study. For each city, we report the average PT score, $PT_{city}$, vs the fraction of the population living in a 15-minute condition, $F_{15}$. The circle size is proportional to the population density of the city. At the same time, the colour codes for the Gini index. This scatter plot features interesting patterns. The triangular grey area represents a theoretically unreachable phase: if a city has a low $PT_{city}$, the percentage of people within 15-minute accessibility cannot be lower than a certain percentage. We observe a decreasing trend of $F_{15}$ vs $PT_{city}$, which is to be expected. Less trivial is that the Gini index features an increasing trend with $PT_{city}$.}
\label{fig:accessibility_scores}
\end{figure*}

We start our analysis by conducting a thorough assessment of the closeness of current cities to the ideal of 15-minuteness. The first significant result, extracted from the study of many cities worldwide, is a pretty substantial heterogeneity of accessibility both within and across different cities. In this manuscript, we will illustrate more than 50 cities in depth, and we refer the reader to the online platform that we freely made available, which now covers around ten thousand cities worldwide. For all cities, we define and measure the so-called Proximity Time,  $\operatorname{PT}$ (see the Methods section for its definition) both locally and for each city as a whole. The Proximity Time measures how long it takes from a specific point in the city to reach the services of the 15-minute city basket, either by walking or biking. Panel A of Fig.~\ref{fig:accessibility_scores} reports the maps of PT values (by walking) for some selected cities. The first striking evidence refers to the remarkable differences among different cities. Equally, remarkable differences are also observed within a single city, implying that accessibility is not a ``currency'' equally distributed in the population, with a centric structure of cities or polycentric one in some notable cases, e.g., Barcelona and Paris.

To quantify the level of inequality in accessibility, we measure, for each city, the fraction of the population living in a 15-minute condition, i.e., the fraction of residents with access to essential services within a 15-minute radius. We name this quantity $F_{15}$. Panels B and C of Fig.~\ref{fig:accessibility_scores} quantify the inequalities in accessibility within and across cities. Two striking results emerge. First, there is a considerable variation in the fraction of the population living in a 15-minute condition across different cities (Panel B). Second (Panel C), within the same city, the variations in accessibility can be significant, and they tend to grow with the average $PT$ score, meaning that cities with average bad accessibility are also more unequal. Finally, substantial variations exist among cities in different cultural contexts, mainly based on how car-centric and suburban is the typical urban planning style.

Our analysis consistently highlights patterns where city centres have better access to services than peripheral areas. However, notable exceptions exist, such as Paris or Barcelona, whose recent policies on increasing local access to services are well known~\cite{pozoukidou202115}. These cities exhibit a more evenly distributed accessibility, transcending the typical centre-periphery divide. It is important to remark that it is in the nature of cities, however, to provide more services in the centres: it is the most convenient place for people from different areas to meet, and therefore an attractive place for offices and shops. But the self-reinforcing process of "rich get richer" has to be avoided: more services mean more people willing or needing to visit the city centres and an ever-increasing demand for services in the centres, making the peripheries ever more empty and isolated.

\subsection{Towards an optimally equal distribution of opportunities}

\begin{figure*}[!ht]
\centering
\includegraphics[width=\textwidth]{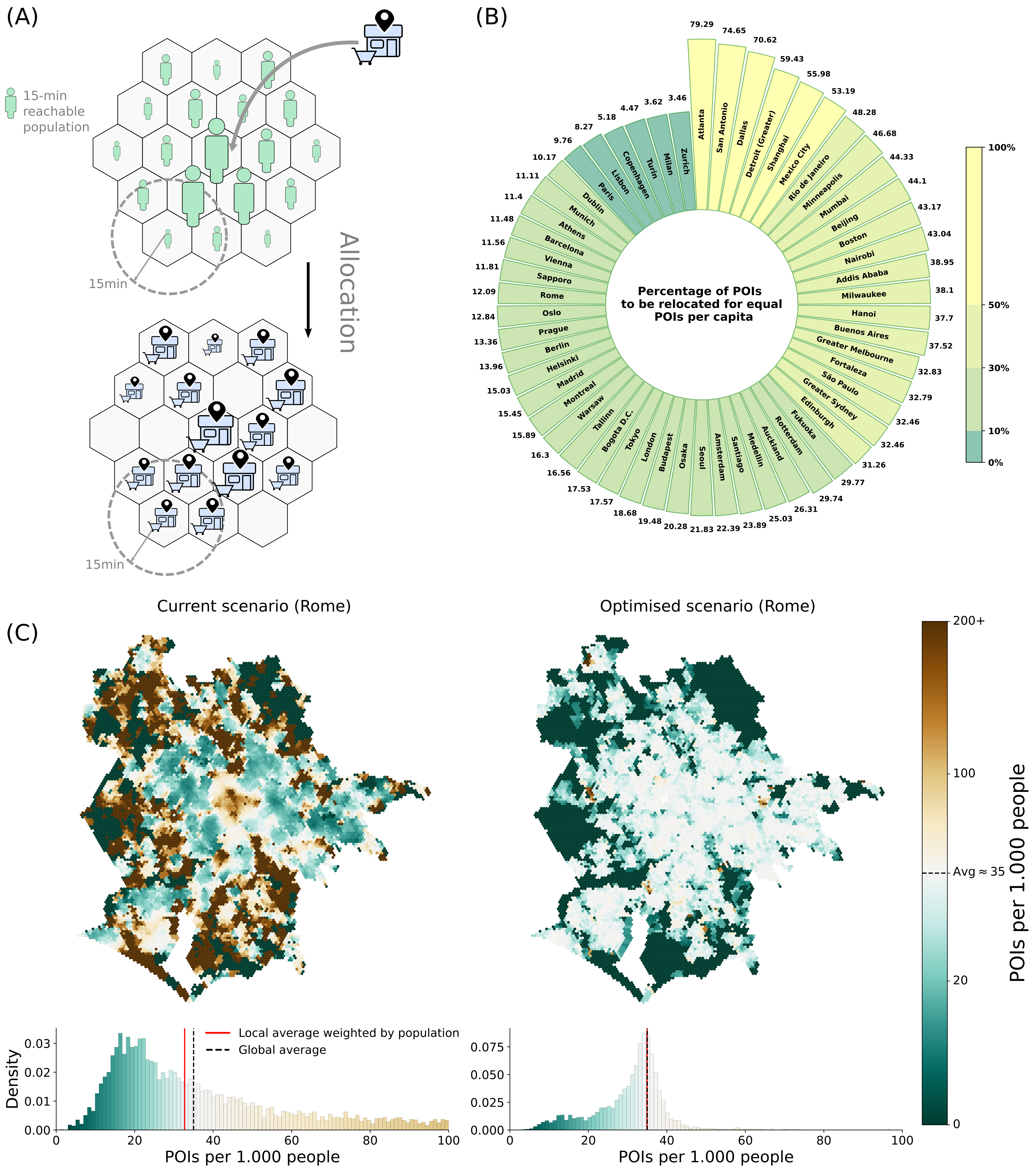}
\caption{\textbf{A scheme of the algorithm and its effects on the POIs distribution.} {\bf Panel A} shows the rationale below the algorithm (we refer to the Methods section for a thorough description): the more people are reachable in 15 minutes from a certain area in the city, the more POIs will be placed there and the surroundings. {\bf Panel B} reports a ranking of the fraction of the POIs relocated POIs in each city after applying the algorithm above. A stark variation across cities is immediately noticeable. {\bf Panel C} depicts two maps of Rome coloured according to the local \textit{number of POIs per 1000 people}, equivalent to POIs per capita, calculated before and after applying the relocation algorithm. Each map is complemented by a histogram of the number of POIs per 1000 people, again before and after using the relocation algorithm. The algorithm makes the local availability of services per capita close to the global average in the city.
}\label{fig:reallocation_algorithm}
\end{figure*}

Having assessed the levels of inequality in urban accessibility, it is pretty natural to raise the question of what could be done to mitigate those inequalities. Lacking systematic empirical examples, we could first ask whether a suitable reallocation of the existing POIs could lead to better accessibility patterns and, on a quantitative ground, to higher values of the fraction of the population living in a 15-minute condition, $F_{15}$. To this end, we have devised an algorithm that optimally redistributes the POIs that already exist in a given city. Panel A of Fig.~\ref{fig:reallocation_algorithm} describes the algorithm's rationale for redistributing POIs based on the population distribution within each city. The algorithm aims to have an equal amount of services per capita, or equivalently per 1000 people, across the city, leading to a more balanced distribution of services than the actual distribution. In this scenario, every amenity will serve approximately an equal amount of people in its neighbourhood.

The application of our algorithm yields some insights when considering how many services need to be relocated in different areas from their original ones. Panel B of Fig.~\ref{fig:reallocation_algorithm} illustrates that cities that are known for their car-centric designs, such as Atlanta and other North American cities, need to relocate a high percentage of POIs, over 70\% in some cases; conversely, certain European cities, including Milan, Copenhagen, Lisbon and Paris, already demonstrate a well-optimised, homogeneous distribution of services. Here, the need for reallocation is minimal, reinforcing the effectiveness of the present urban design in making the city walkable and accessible. These findings highlight the diverse challenges different cities can face as they strive to improve their accessibility and become 15-minute cities.  Panel C reports an example of how the relocation algorithm changes the map of the local number of POIs per 1000 people in Rome.

\begin{figure}[h]
\centering
\includegraphics[width=\columnwidth]{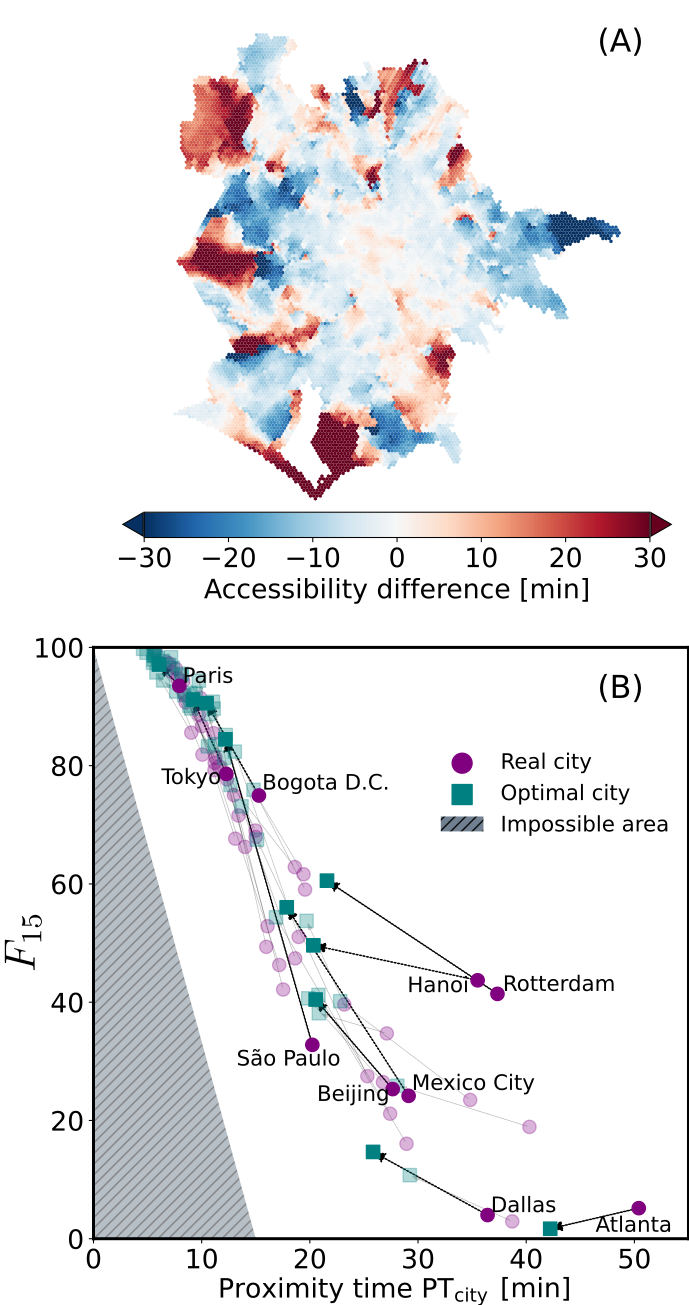}    
\caption{\textbf{Impact of POIs relocation on accessibility} {\bf Top} panel shows the accessibility difference, in minutes, after applying the algorithm to the city of Rome: red areas get worse because they are over-served or not inhabited. In contrast, blue areas are under-served and consequently improve their accessibility. {\bf Bottom} panel shows the change in the measures of local accessibility depicted in Fig.~\ref{fig:accessibility_scores} (Panel C) after applying the algorithm.}
\label{fig:accessibility_improvements}
\end{figure}

Let us see now how the reallocation of POIs affects accessibility across various cities. We present these shifts in accessibility in Fig.~\ref{fig:accessibility_improvements} through two distinct visuals. In particular, panel A represents a map of Rome with a colour code associated with the changes in accessibility (i.e., in the $PT$ score) at the local level. By redistributing POIs, we can appreciate areas of significant positive changes (blueish areas), substantial negative changes (reddish areas), and substantially unchanged areas. It is remarkable how the city centre does not get depauperised of services despite the growth in accessibility of other less central areas. Further, the peripheral regions experience a substantial enhancement in their access to services, validating the efficiency of the reallocation strategy in bridging accessibility disparities. Reddish areas get worse in accessibility, presumably because they were already well served or more likely because they are less densely populated.

Panel B of Fig.~\ref{fig:accessibility_improvements} reports a scatter plot in the plane $F_{15}$ vs $PT_{city}$, analogous to that of Fig.~\ref{fig:accessibility_scores} (Panel C). Here, the arrows illustrate the changes a selected set of cities undergo. It is evident how the redistribution of the existing POIs triggers a leftward and upward stream of the cities, i.e., a trend towards lower values of $PT_{city}$ (larger average accessibility) and higher $F_{15}$ (larger equality). A notable exception is represented by Atlanta, which deserves a deeper insight. Given the high level of sprawl in Atlanta, the few compact neighbourhoods in the centre of Atlanta worsen their accessibility. Thus, the city sees a small decline in the percentage of residents within 15-minute areas, despite witnessing an improvement in its average local accessibility.

Our results emphasise the hypothesis that the '15-minute city' paradigm is suitable for relatively compact (i.e., high-density) urban areas. Even if a strategic reallocation of services or an increase of their capillary penetration can notably improve accessibility, the whole paradigm of the '15-minute city' has to be rethought in areas with lower urban densities, such as suburban or intermediate zones.  We shall come back to this point later on.

\subsection{The quest for inclusive 15-minute cities}

So far, we have explored the possibility of better distributing the available opportunities in a city to improve the equality of access to them. Now, we take a step further and ask ourselves about the optimal number and distribution of POIs to make a city 15-minute for the largest population fraction.

We can employ our relocation algorithm to simulate any quantity of POIs in a given city. In particular, we are interested in quantifying the number of POIs required to become a 15-minute city. With this aim in mind, we simulated the optimal city structure considering an increasing number of POIs. We ran our algorithm independently for each POI type (all services are treated equally in our study). This approach allows us to introduce a parameter for gauging a city's accessibility: the number of POIs per 1000 people.

\begin{figure}[h]
\centering
\includegraphics[width=\columnwidth]{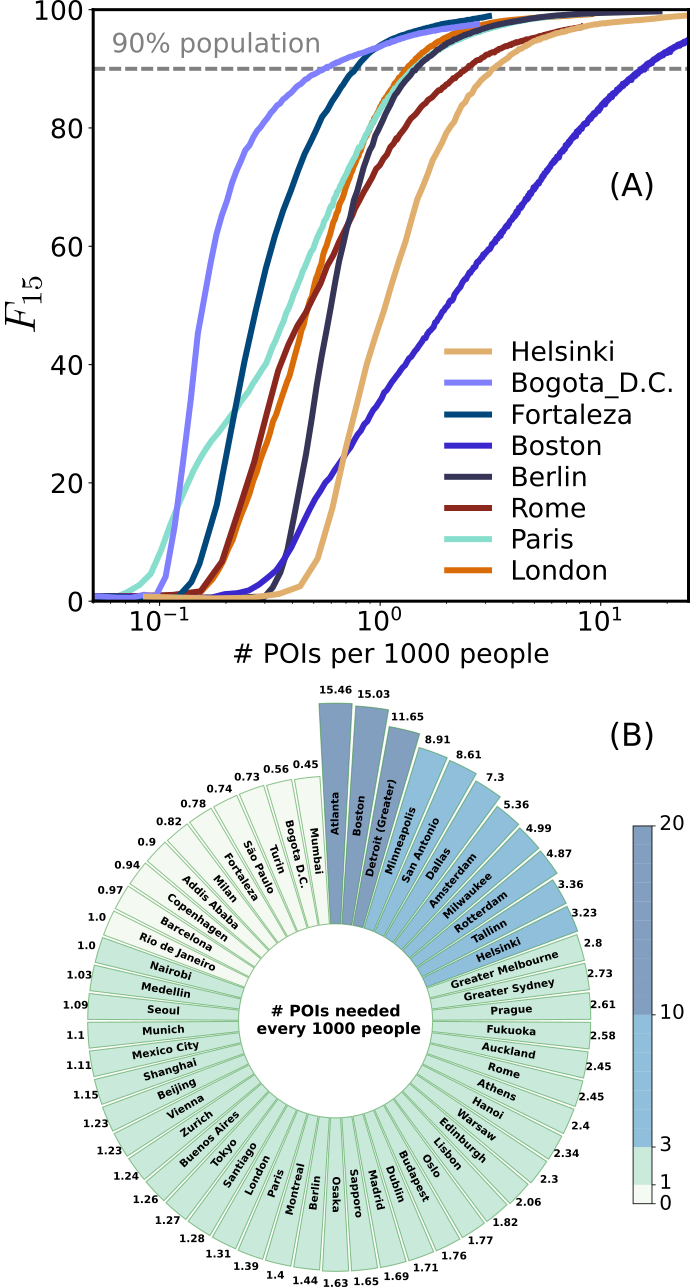}
\caption{\textbf{Optimal number of POIs per 1000 people.} {\bf Top}: values of $F_{15}$ as a function of an increasing number of POIs per 1000 people distributed in the city by our algorithm. The intersection with $F_{15}=90\%$ defines the number of POIs per 1000 people needed to become a 15-minute city. {\bf Bottom}: radial histogram reporting the number of POIs per 1000 people needed to become a 15-minute city. This number is extremely heterogeneous, resulting from very diverse local and global population densities.}
\label{fig:optimal_curves}
\end{figure}

In Fig.~\ref{fig:optimal_curves}, we present the results of our simulations. Panel A presents the percentage of the population with a proximity time below 15 minutes, $F_{15}$, which features strong variation across cities. We compute, in particular, the intercept of the curves with the horizontal line $F_{15}=90\%$, i.e., the necessary number of POIs per 1000 residents for 90\% of the city's population to live within a 15-minute access area. Note that this number is not the bare minimum required to have widespread 15-minute accessibility, since the hypothesis of a homogeneous density of POIs per capita is not always satisfied. 

The quantity of POIs per capita required for a city to be a '90\% 15-minute city' displays considerable variation among the cities studied. With reference to Panel B of Fig.~\ref{fig:optimal_curves}, for instance, highly dense and homogeneous cities, like Mumbai and Bogota, require only 0.45 and 0.56 POI per 1000 residents, respectively,  to achieve the 15-minute city. Similarly, it occurs in many other South American cities such as S\~{a}o Paulo, Mexico City and Fortaleza. Of course, the debate is wide open about the potential overcrowding of POIs (and for sure about their quality), but this topic falls beyond the scope of our current study.

European and Asian cities present a varied landscape. They lie in an intermediate zone, displaying differing trends based on their urban layout and design. For instance, cities historically designed with a greater focus on automobile transit, like Rotterdam, or with large suburban areas, like Tallinn, Edinburgh, Amsterdam or Helsinki, require a higher number of POIs per capita to become '15-minute cities'. This is also the case in cities in Australia and New Zealand. In contrast, compact cities, such as Turin, Milan and Barcelona, demand fewer POIs per capita due to their dense and homogeneous urban structure. Another peculiar case is that of cities with discontinuous urban textures, such as Rome or Athens, for which historical heritage and geographical constraints represent an additional challenge towards a compact design. 

As already observed, many US cities stand as outliers and represent unique challenges. To achieve 15-minute city status, these cities would require an excessively high number of POIs. Atlanta, for example, would need more than 15 POIs per 1000 residents, which converts to one POI for every 64 residents. Such a requirement appears unsustainable, once again emphasising the impact of city density and design on the feasibility of the 15-minute city concept.

\section{Methods} 

\subsection{Data Acquisition}

The data acquisition stage involved collecting four core datasets:

\begin{itemize}
\item \textbf{City boundaries:} the boundaries of the cities were acquired from the Organisation for Economic Co-operation and Development (OECD) shapefiles~\cite{dijkstra2019eu}, focusing on the defined core city. When OECD data were unavailable, we used the Global Human Settlement (GHS) files~\cite{florczyk2019ghs}, focusing on the city's core for the analysis.

\item \textbf{Points of Interest (POIs):} amenities, which we will call generically POIs, were sourced from OpenStreetMap~\cite{OpenStreetMap} and classified according to their tags. Only POIs representing services for people were retained, and non-service POIs such as trees and buildings were disregarded.

\item \textbf{Population data:} Population data was derived from WorldPop~\cite{worldpop}, providing demographic context for our accessibility measures. Our study used a 100m population density grid corrected to match municipal UN population estimates ~\cite{bondarenko2020census}.

\item \textbf{Times of accessibility:} The times of accessibility to POIs from points in the city were calculated using Open Source Routing
Machine (OSRM)~\cite{luxen2011osrm}, based on OpenStreetMap data.
\end{itemize}

\subsection{Data Preparation}

The data preparation stage encompassed two main steps:

\begin{itemize}
\item \textbf{Hexagonal grid generation:} A regular hexagonal grid with a side length of 200 meters was generated across the city. Hexagons without any nearby POIs were omitted from further analysis.

\item \textbf{POI categorisation:} The retained POIs were categorised into nine service types: outdoor activities, learning, supplies, eating, moving, cultural activities, physical exercise, services, and healthcare. This categorisation allowed for both general and category-specific accessibility assessment.
\end{itemize}

\subsection{Accessibility Calculation}

With the prepared data, we proceeded to compute the accessibility. Our measure of accessibility is similar to what has been proposed in the literature with the name of \textit{dual access}~\cite{cui2020primal}. We compute it as follows:

\begin{itemize}
\item \textbf{Hexagon-level accessibility:} For each hexagon $k$, we identified the 20 nearest POIs in each category using OSRM for both walking and biking routes. In the main text, we only show the results for the simulations for accessibility by walking, and we refer to the online platform for the simulations about accessibility by bike. The average time $\langle t \rangle_{c,k}$ required to access the 20 nearest POIs of category $c$ in the hexagon $k$, is computed as follows:

\begin{equation}
\langle t \rangle_{c,k} = \frac{1}{n} \sum_{i=1}^{n} t_i^{c,k}
\end{equation}

\noindent where $n$ is the number of POIs (in this case, $n=20$) and $t_i^{c,k}$ is the time required to reach the $i$-th POI of category $c$ in the hexagon $k$. This procedure yields the accessibility value for that category in that hexagon, and it is equivalent to calculating the dual access to the 20 closest POIs for each hexagon.

\item \textbf{City-level accessibility:} The accessibility measures for each hexagon $k$ were averaged to generate a single proximity time index $\operatorname{PT}_k$ for that hexagon:

\begin{equation}
\operatorname{PT}_k = \frac{1}{m} \sum_{c=1}^{m} \langle t \rangle_{c,k}
\end{equation}

\noindent where $m$ is the number of categories. These indices were then averaged across the city, weighted by the population $p_k$ within each hexagon, to compute the overall city accessibility $\operatorname{PT_{city}}$:

\begin{equation}
\operatorname{PT_{city}} = \frac{\sum_{k=1}^{K} \operatorname{PT}_k p_k}{\sum_{k=1}^{K} p_k}
\end{equation}

%\noindent For simplicity, we will call the city-wide proximity time $\operatorname{PT_{city}}$ as just $\operatorname{PT}$.  

\noindent where $K$ is the number of hexagons, $\operatorname{PT}_k$ is the proximity time index of the $k$-th hexagon, and $p_k$ is the population of the $k$-th hexagon. The corresponding category measures were averaged to derive specific city scores for individual categories.

\item \textbf{15-minute accessibility:} We also computed the percentage of the city's population within a 15-minute access range, providing another measure of city-wide accessibility:
\begin{equation}
F_{15} = \frac{\sum_{k=1, PT_k \leq 15}^{K} p_k}{\sum_{k=1}^{K} p_k} \times 100
\end{equation}
\end{itemize}

\subsection{Inequality Assessment}

Lastly, to measure the level of inequality in accessibility across the city, we calculated the Gini inequality index $G$:
\begin{equation}
G = 1 - \frac{2 \sum_{p=1}^{\operatorname{N_{pop}}}\sum_{p'\leq p} \operatorname{PT}_{p'}}{\operatorname{N_{pop}}\sum_{p=1}^{\operatorname{N_{pop}}} \operatorname{PT}_p}
\end{equation}
\noindent where $\operatorname{PT}_p$ is the $p$-th proximity time measure \textit{of a single person} in the city when sorted in non-decreasing order.

\noindent This metric provides a non-trivial measure of urban accessibility, revealing how unequal local accessibility is in the city. Anyway, a compact city with a good proximity time will not necessarily have a low Gini coefficient because the accessibility times can still fluctuate.

\subsection{POI Reallocation Algorithm}

Our algorithm takes the population distribution within the city's hexagonal grid as input and optimises the allocation of POIs independently of the category. The optimisation is performed separately for each category. Let us briefly describe the rationale of the algorithm, whose steps are depicted in Figure \ref{fig:reallocation_algorithm}. The optimisation aims to keep the number of POIs per capita constant in the city to serve every area equally, considering its population density. Therefore, each POI will have to serve on average the same number of people, which we call \textit{capacity} $\operatorname{CAP}$ and is a global quantity of the considered city and category:
\begin{equation}\label{eq:pois_capacity}
    \operatorname{CAP}_c = \frac{N_{\text{pop}}}{N_{\text{c}}} \: \text{,}
\end{equation}
\noindent where $N_{pop}$ is the total population of a city, $c$ a category of services and $N_c$ is the number of POIs of category $c$ in the city.

The inverse of the capacity is the number of POIs per capita, reading:

\begin{equation}\label{eq:pois_per_capita}
    P_{\text{c}} = \frac{N_{\text{c}}}{N_{\text{pop}}} \: \text{.}
\end{equation}

Note that we use the number of POIs per thousand people $1000 \cdot P_c$ instead of POIs per capita to improve the measure's readability and meaning.

A city is initially considered empty and then iteratively filled with POIs. Each iteration places a POI in the hexagon with the highest population demand (i.e., population needing a POI) that can be reached within a chosen time threshold, 15 minutes in our case. Actually, we make the algorithm slightly random by randomly choosing a neighbour of the highest demand hexagon to avoid the effect of always choosing the cells with the locally highest needing population. After placing the POI, a number of people equal to the POI capacity $\operatorname{CAP}_c$ will be deducted from the needing people in the hexagons that can reach the POI in 15 minutes, in amounts proportional to the population demands of the hexagons. At the end of this iteration, the total sum of the needing population is reduced by exactly $\operatorname{CAP}_c$. This operation is iterated until all POIs have been assigned to a hexagon, yielding an equal distribution of services per capita across the city. The algorithm is the same for all categories, and it is dependent only on the city's population distribution. The number of POIs per capita in each 15-minute neighbourhood should then be (roughly) the same across the city and equal to the average POIs per capita in the city (see Figure \ref{fig:reallocation_algorithm}). 
A pseudocode for the algorithm can be found in the supplementary material.

\section{Discussion and conclusions}

This study contributes to the ongoing conversation on urban planning and design by quantifying local accessibility worldwide and offering an empirical method for simulating scenarios towards proximity-based cities that optimise equal access to opportunities. The contribution of this paper is twofold: methodological and conceptual.

On the methodological side, we originally introduced a tool to quantify how close a city is to the ideal of being 15-minute. We adopted this metric to assess the status of many cities worldwide. We provided an open-access platform (\url{whatif.sonycsl.it/15mincity}) for everyone to explore cities or portions of them. The second crucial methodological contribution is the conception and implementation of a heuristic algorithm to redistribute Points of Interest in a city based on giving equal accessibility to the largest population fraction. Although relocating activities can be a slow process, POIs related to commercial activities represent a very fast-paced environment where many activities can be displaced in a relatively short time, looking for better opportunities. In addition, in many countries worldwide, the average building lifespan starts to be smaller than human lifespan expectancy. We tested the algorithm in two conditions: the relocation of existing POIs and the quest for optimal POIs given the local population densities. 

On the conceptual side, this paper provides several essential observations. 

First, we provided a global picture of current local accessibility worldwide. We draw accessibility maps based on the computation of a brand-new observable, the Proximity Time score ($PT$), that quantifies the time needed from each local area to reach the essential urban services by walking and biking. Our results highlight a profound heterogeneity in accessibility patterns within and across cities, mirroring different aspects, from geomorphological/ecological to historical/cultural and management. In the comparison among cities, marked variations appear, with only a tiny part of the cities very well positioned to be 15 minutes and an extensive distribution on more or less large values of the score. Stark contrasts are also found in different areas of the same city, with the inequality of accessibility usually following a core-periphery gradient and a few cities that seem to have more polycentric accessibility to services than others. The disparity in the local availability of services is an additional layer of difference between cities, as some, despite an excellent average local accessibility, are highly heterogeneous.

Second, we asked ourselves about the possibility of improving present cities, considering the equality of access to urban opportunities. The first question has been whether a better geographical redistribution of the existing POIs would enhance the accessibility patterns for a more significant fraction of the population. We ran this simulation through our original relocation algorithm for many different cities, and we concluded that already relocating the existing POIs (i.e., without deploying new ones) achieves the objective of significantly improving the proximity of peripheral and underserved areas while not impoverishing the central, already well-served, areas. This result advocates for more decentralisation policies of the cities' activities, favouring populated peripheral areas.

Finally, we asked whether it is possible to conceive cities that would be optimal from the perspective of equal access to opportunities. In this case, we simulate scenarios of the geographical distribution of POIs by removing the constraints of the present distribution of POIs. Instead, we simulate the best positioning of an increasing number of POIs starting from an empty city (i.e., without POIs). This simulation allows us to identify, for each city, the minimum number of POIs needed for the city to be 15-minute for at least 90\% of their residents. This observable nicely illustrates how hard it is for a city to become 15-minute, and how large the fluctuations are across cities. In some urban contexts, particularly in US cities, the number of POIs per capita would be too large to be economically sustainable. A scenario emerges in which the very notion of the 15-minute city can not be a one-fits-all solution and is not a viable option in areas with a too-low density and a pronounced sprawl. On the other side of the spectrum, in many dense cities, the number of POIs needed is relatively low and can thus be achieved with minimum effort from policy-makers. Paris, Barcelona, and Milan represent compelling examples demonstrating the efficacy of urban planning policies oriented towards local accessibility. These examples are potent arguments for the feasibility of 15-minute cities in dense urban areas and offer valuable lessons for other cities aiming to achieve similar outcomes. As for more sprawl cities, for instance, US cities, the low densities doom them far from the idea of a proximity-based city. 

Since a substantial increase in the population density in already inhabited areas is not an option in many concrete cases, a more general theoretical framework~\cite{logan2022x} is in order to include local population densities as a relevant variable for non-compact urban contexts. Only some of the POIs have the same degree of replication. One can easily open a new grocery store virtually everywhere, but one cannot replicate a landmark monument or a significant infrastructure (hospital, opera theatre, football stadium) in every neighbourhood. For this reason, it is interesting to promote a slightly different notion of city in which what is optimised is not only the transit time but the amount of opportunities, wherever they are, available to residents. 

Before concluding, let us remark that since our analysis is based on open data, it can be subject to some limitations depending on possible sources of bias in the data. For instance, POIs data coverage can be limited in some areas or cities, or the walkability of some areas could be different in some cases~\cite{rhoads2023inclusive}. These minor issues can be progressively solved using more precise and complete data. The interested reader can find a more detailed discussion about possible critical points in the Supplementary Information.
% For instance, OpenStreetmap data can cover more central areas of cities or more developed countries~\cite{herfort2023spatio}, whose services are easier to map, and the times of accessibility can be underestimated in some cases due to lack of information about the conditions of the infrastructures~\cite{rhoads2023inclusive}. Similar known biases can come from the world population grids or the definition of urban areas. The interested reader can find a more detailed discussion about possible critical points in the supplementary material.

In conclusion, our research represents a significant step forward in understanding urban accessibility and the viability of proximity-based cities. While our findings might provide a new perspective on traditional planning approaches, they offer practical and actionable insights for designing sustainable, accessible, and liveable urban environments. Our work also highlights that the ideal of a proximity-based city heralded by the 15-minute city must be balanced with faster and more reliable public transportation services that allow for eliminating some of the disparities in peripheries, particularly those due to the heterogeneity of local population densities. Finally, a broader perspective integrating socio-economic and cultural factors should be included as a next step to shift focus from only time-based to value-based cities. As we walk towards imagining future cities, we hope our research catalyses deeper exploration and triggers concrete actions in urban planning and policy.

\backmatter

\bmhead{Supplementary information}

Supplementary information is available in the attached file.

\bmhead{Competing interests}
The authors declare no competing interests.
\bmhead{Authors' contributions}
All authors designed the research and analysis. BC and MB performed the data analysis. VL, MB and HPMM wrote the manuscript. All authors revised the manuscript.
\bmhead{Acknowledgments}
The authors thank Bernardo Monechi and Claudio Chiappetta for their contributions.

\bibliography{biblio}

\clearpage 
\newpage

\begin{appendices}

\onecolumn
% \appendix

\begin{center}
    \Large{\textbf{Appendix}}
\end{center}
% \section{Section title of first appendix}\label{secA1}

\section{The proximity time scores and rankings}
\begin{figure*}[!htb]\label{fig:si_time_rankings}
\includegraphics[width=.49\textwidth]{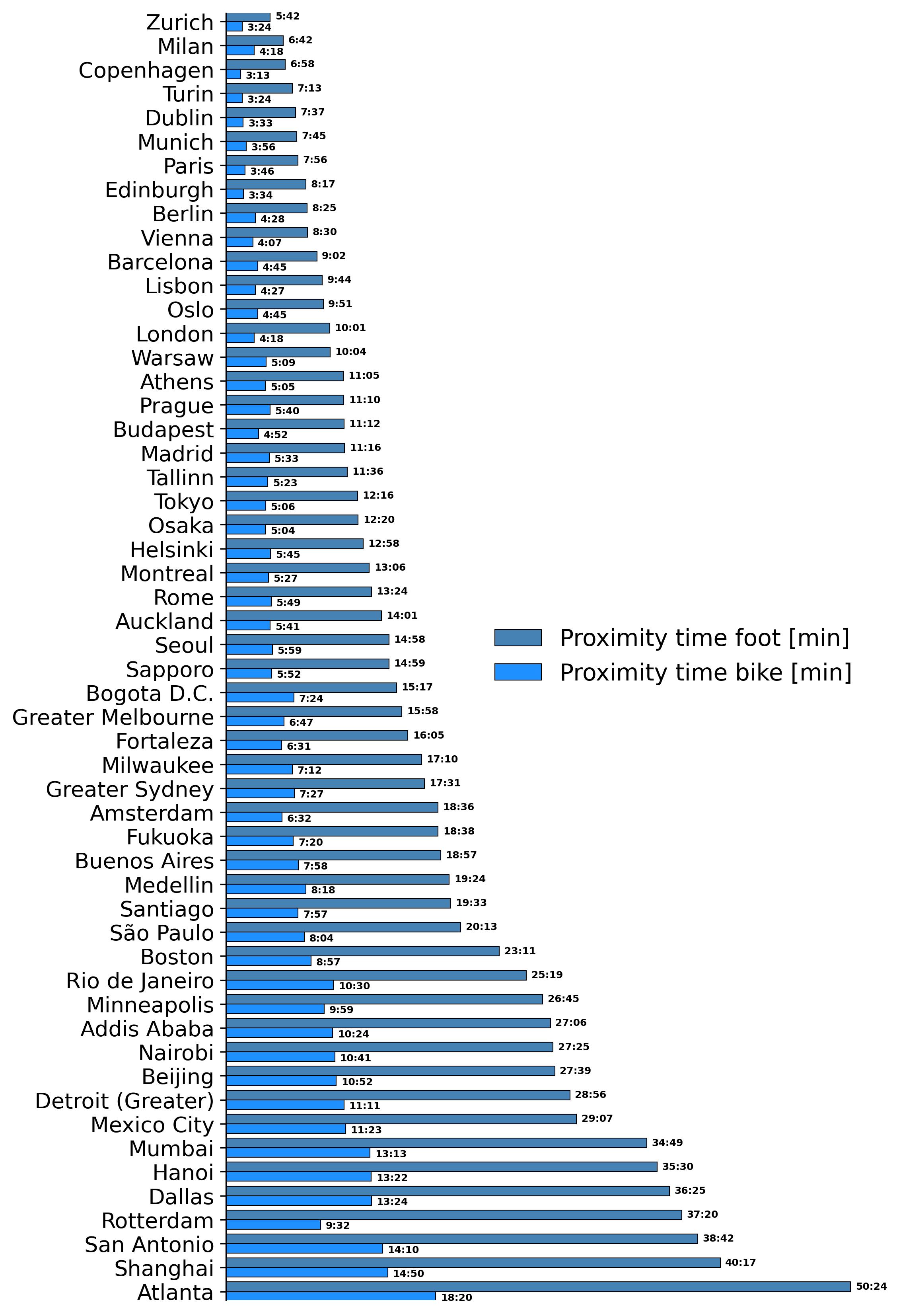}
\includegraphics[width=.49\textwidth]{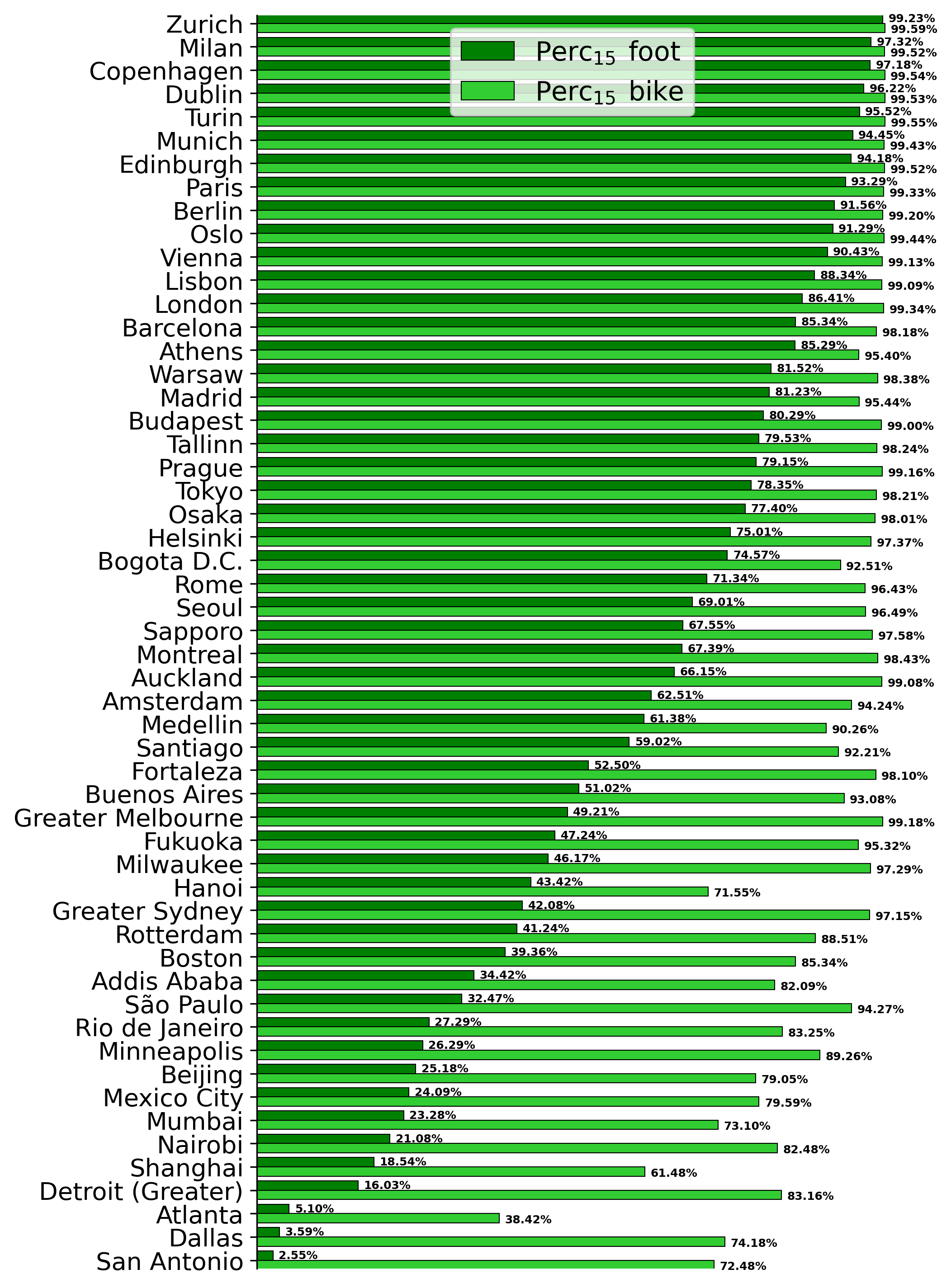}
\caption{\textbf{The rankings of accessibility.} The left plot describes the rankings of the average Proximity Time of the cities, both by bike and on foot, while the right one describes the percentages of people in a 15-minute city. } 
\end{figure*}

We present here a closer look at the accessibility scores for the set of cities we considered in the main text. In Figure \ref{fig:si_time_rankings}, we present both the city average and the percentage of people within 15-minute accessibility on foot and by bike. The left plot ranks the cities according to their proximity time by foot. It is also interesting to compare the foot and bike accessibility, which are close or far depending somehow on how bikable the city is: for instance, Rotterdam and Amsterdam have very good bike accessibility compared to the walking one, whereas cities like Hanoi and Mumbai or Bogotá show the opposite trend. The right plot shows the percentage of people living in 15-minute areas again by bike and on foot, and the same considerations are valid. In both cases, it is remarkable to see that most cities would be close to the 15-minute city if they were free from car traffic and therefore more suitable for bikes: our analysis does explore how easy it is to bike or walk in cities (see SI \ref{si:subsec:bias}).

\section{Sources of bias from the data}\label{si:subsec:bias}

A few remarks are due about the potential limitations of our study. 
\begin{itemize}
    
\item The first critical point is related to the use of OpenStreetMap data. OSM data might not be complete for some cities, and their level of completeness is city-dependent, a long-known issue in OSM data analysis, as highlighted by some recent work~\cite{herfort2023spatio}. Moreover, its open-source categorisation can lead to errors such as repeated entries or wrong categories. It's worth remarking that some parts of our analysis, specifically the simulations of the optimal relocation of POIs, do not depend on POIs data. 
\item The walking infrastructure present in OSM can also lead to biases in times of accessibility to services. Cities might be walkable in principle but less in reality: some areas might be dangerous because of traffic or lack of safety, or the street could be damaged or uphill, therefore not encouraging walkability. In recent work, Rhoads et al.~\cite{rhoads2023inclusive} show that sidewalks can also shape walking trips. This also applies to the biking infrastructure, even more so because it is generally rarer to create a safe, reliable biking network.
\item Another possible source of bias is the definition of urban areas. Here, we considered OECD-defined core urban areas, which rely also on the municipalities' borders\cite{dijkstra2019eu}. When cities are not in the OECD data set, we use borders of core urban areas from GHS instead, but there might be similar biases. However, in most cases, we weigh our accessibility measures by population, meaning that if many empty areas are included in the city shapefile, they will not change the results. 
\item The population data is taken from WorldPop and is crucial for this work. However, it is often considered reliable, and we used data adjusted to match UN population estimates. 
\item The last two points are due to our analysis and can be refined, although it is not trivial to find an unbiased procedure: 
\begin{itemize}
\item First, the categories we chose are arbitrary, and the treatment of all categories in this work is the same, meaning that different services, such as eating and learning (overarching categories including restaurants and schools, respectively), are treated equally despite their specificities. 
\item Lastly, the arbitrary choice of the number of POIs in proximity for the accessibility measure, which is again equal for all services, might lead to some debatable results. The same analysis considering the closest 20 POIs or 2 POIs might change some measures. 
\end{itemize}
\end{itemize}

Future research should address these limitations by considering the differential usage of POIs and improving the data quality to reflect city structures better.

\section{The POIs relocation algorithm}

Here, we present the algorithm with the detailed steps of our methodology \ref{algorithm:pois}. The rationale behind the algorithm is to equalise the number of services per capita across the city. The algorithm is iterative, placing one POI in each iteration and starting from an empty city. It is also category-independent, meaning that the relocation is the same for any kind of service. The random part of the algorithm, yielding slightly different results in different runs, is when the POI is placed in a neighbour hexagon of the "highest demand" cell (step 10 of Algorithm \ref{algorithm:pois}). This choice prevents the algorithm from always choosing a "local maximum" demand cell: if the population needing a service has a local maximum in a certain cell, since we reduce the need for a service homogeneously across the neighbouring hexagons if we always placed the POIs in the highest demand hexagon the local maximum demand cell would not change.

\begin{figure*}[!htb]\label{fig:si_algorithm}
\centering
\includegraphics[width=.34\textwidth]{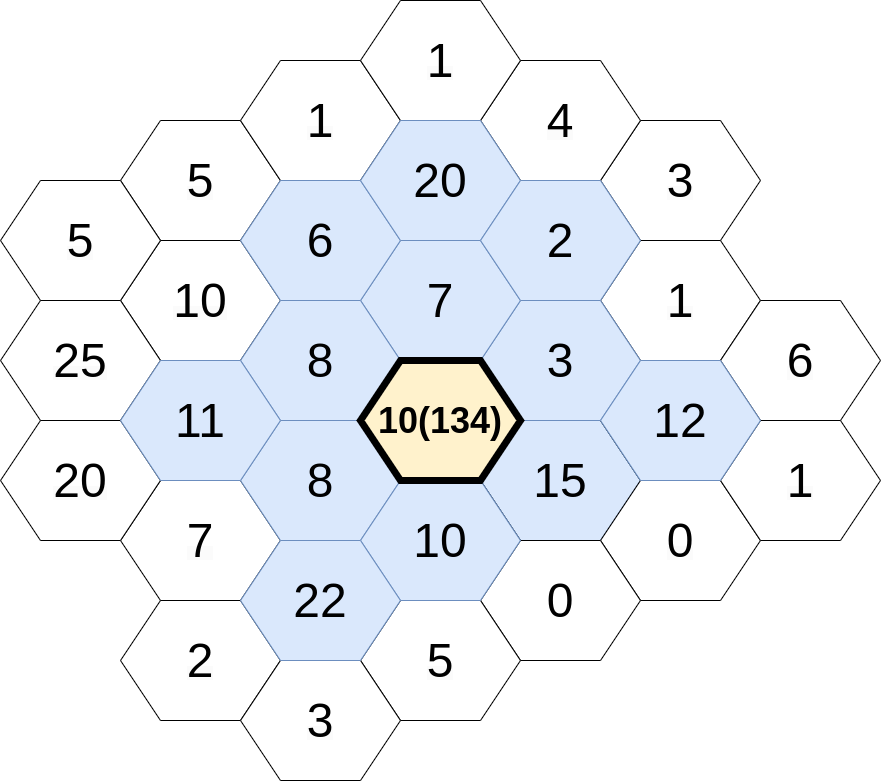}
\includegraphics[width=.34\textwidth]{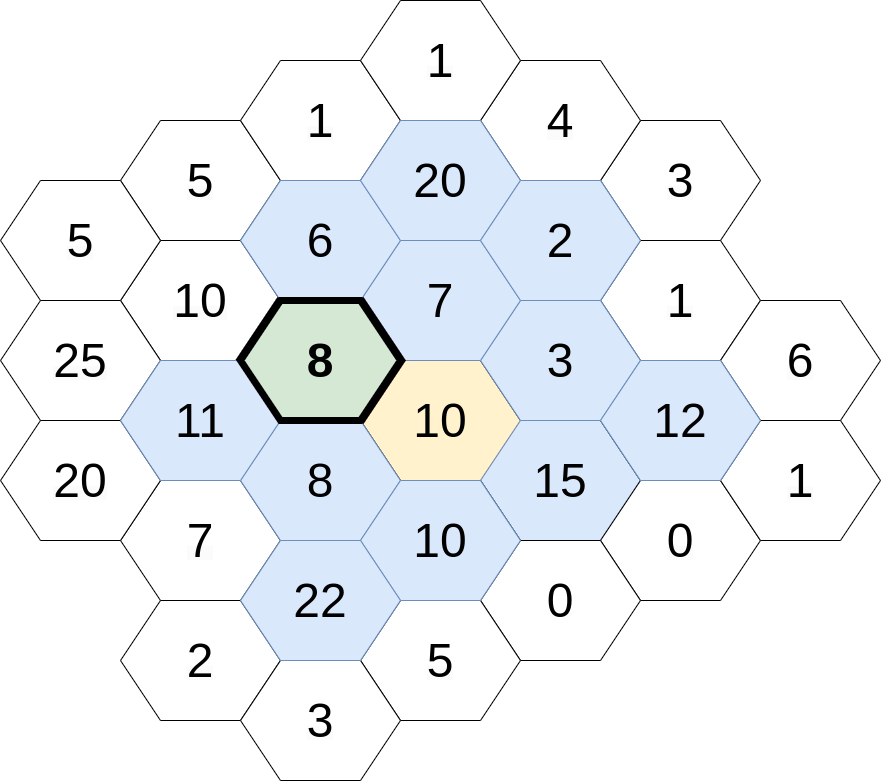}
\includegraphics[width=.34\textwidth]{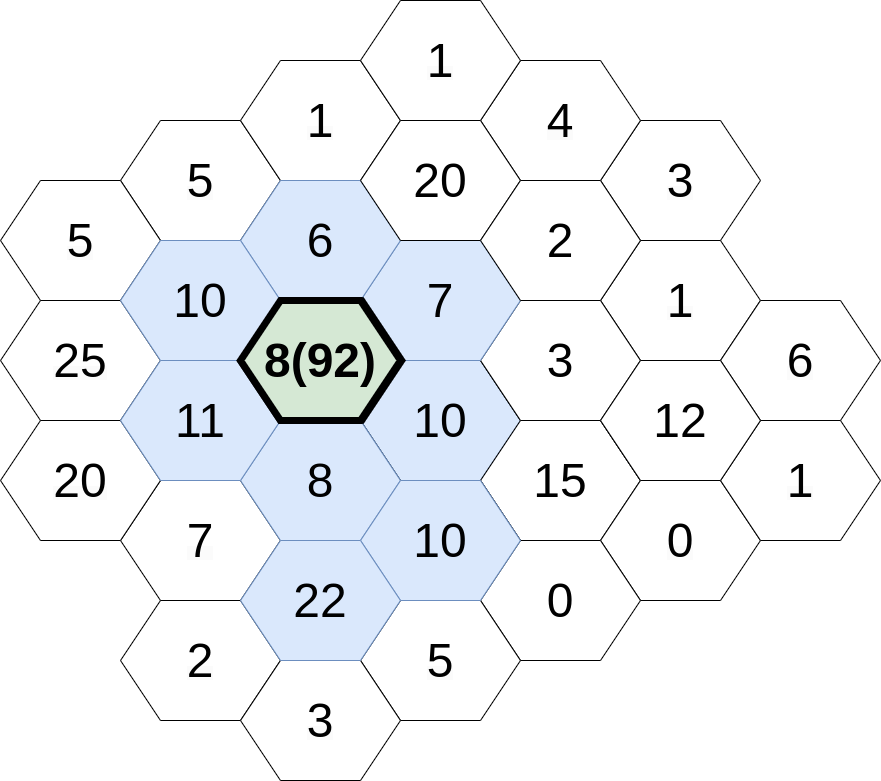}
\includegraphics[width=.34\textwidth]{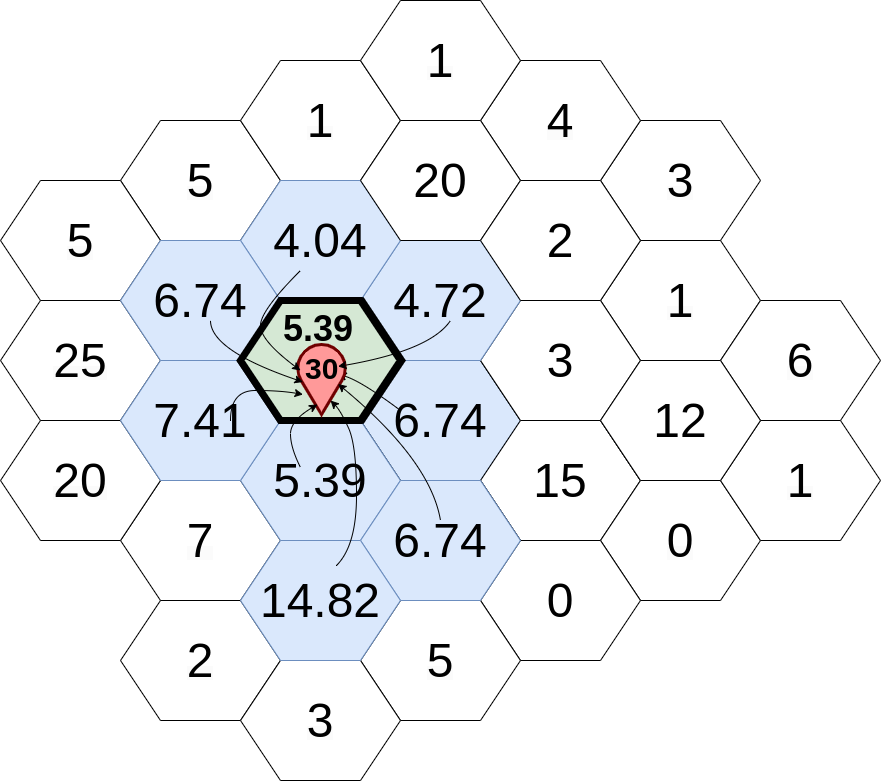}
\caption{\textbf{The POIs relocation steps.} The numbers in the hexagons represent the number of people living in that cell who need a POI, and the numbers within brackets represent the total number of people needing a service in the 15-minute neighbourhood (depicted in blue). The first step selects the hexagon with the largest need for a service in its neighbourhood. Then, a POI is randomly allocated in one of the hexagons of the neighbourhood of the previously selected one. The neighbourhood of this newly chosen hexagon is then calculated, and the needs of the neighbourhood are adjusted so that the number of people served from the allocated POI matches the POI capacity (30 in this case) and that the people needing are reduced proportionally to the resident population.} 
\end{figure*}

We also provide a pictorial representation of the steps of the algorithm in Figure \ref{fig:si_algorithm}.

\begin{algorithm}[!htb]\label{algorithm:pois}
\caption{POI Reallocation Algorithm}
\begin{algorithmic}[1]
% \State Calculate the ratio of population in each hexagon to the total city population;
% \State Determine the number of POIs within each hexagon for the given category;
\State Create a grid for the area of the city;
\State Calculate the total population of each grid cell;
\State Find the 15-minute \textit{neighbourhood} of each cell by finding the cells that can be reached in 15 minutes;
\State Compute the \textit{reachable population} of each cell by summing up the populations of the neighbourhood;
\For {each category}:
\State Compute $\operatorname{CAP}_c$ as the capacity of a POI of the category;
\State Initialise the number of optimally allocated POIs to zero;
\For {each POI in the category}
\State Identify the cell with the highest reachable population;
\State Randomly allocate the POI to a cell within the neighbourhood, with a probability proportional to the resident population;
\State Remove the population from the cells in the neighbourhood that can reach the POI, summing
up to the capacity of the POI and proportionally to the population resident in the cells;
\EndFor
\EndFor
\end{algorithmic}
\end{algorithm}

\clearpage

\section{Accessibility improvements}

\begin{figure}[!htb]
    \centering
    \includegraphics[width=.49\textwidth]{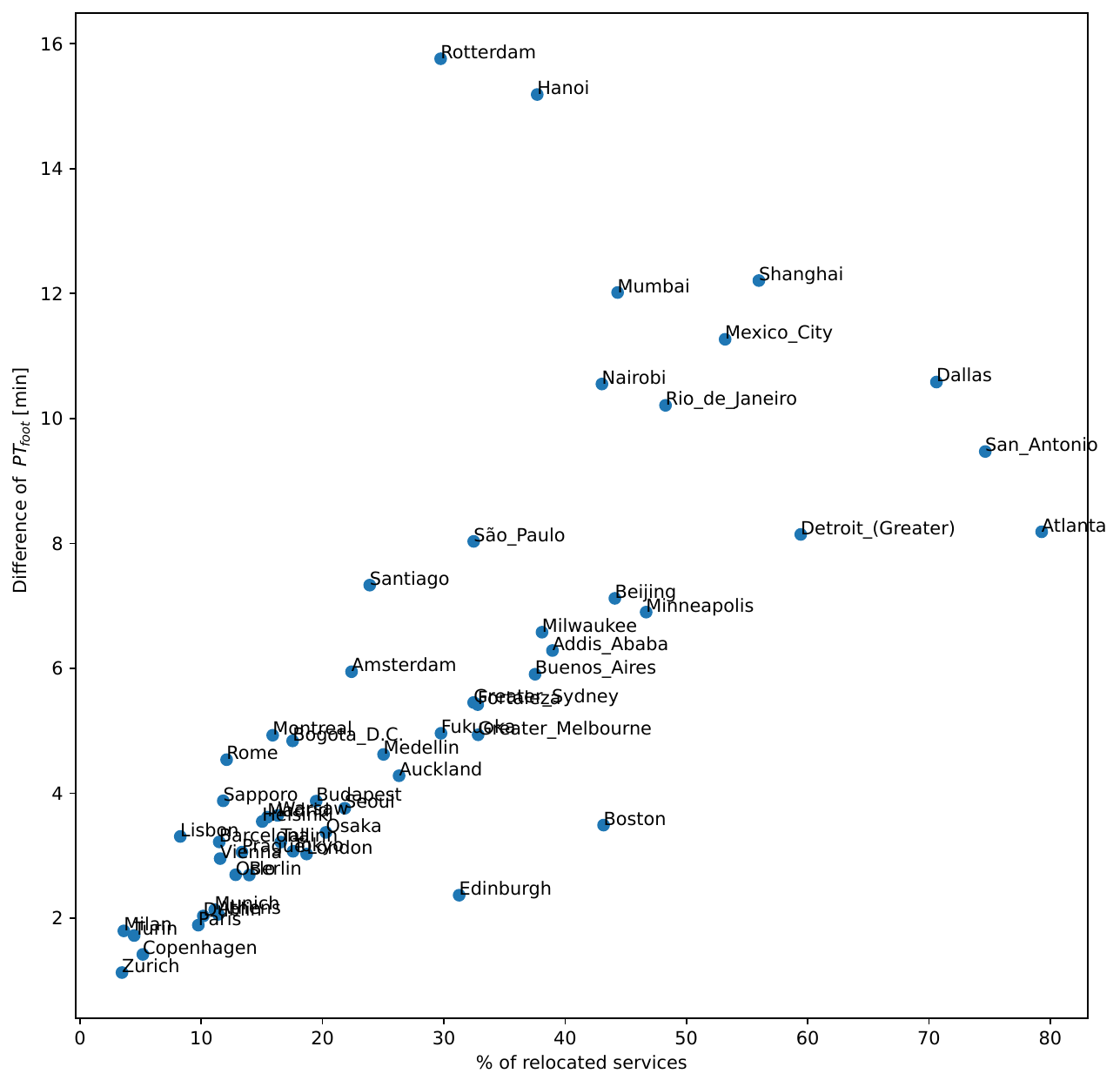}
    \includegraphics[width=.49\textwidth]{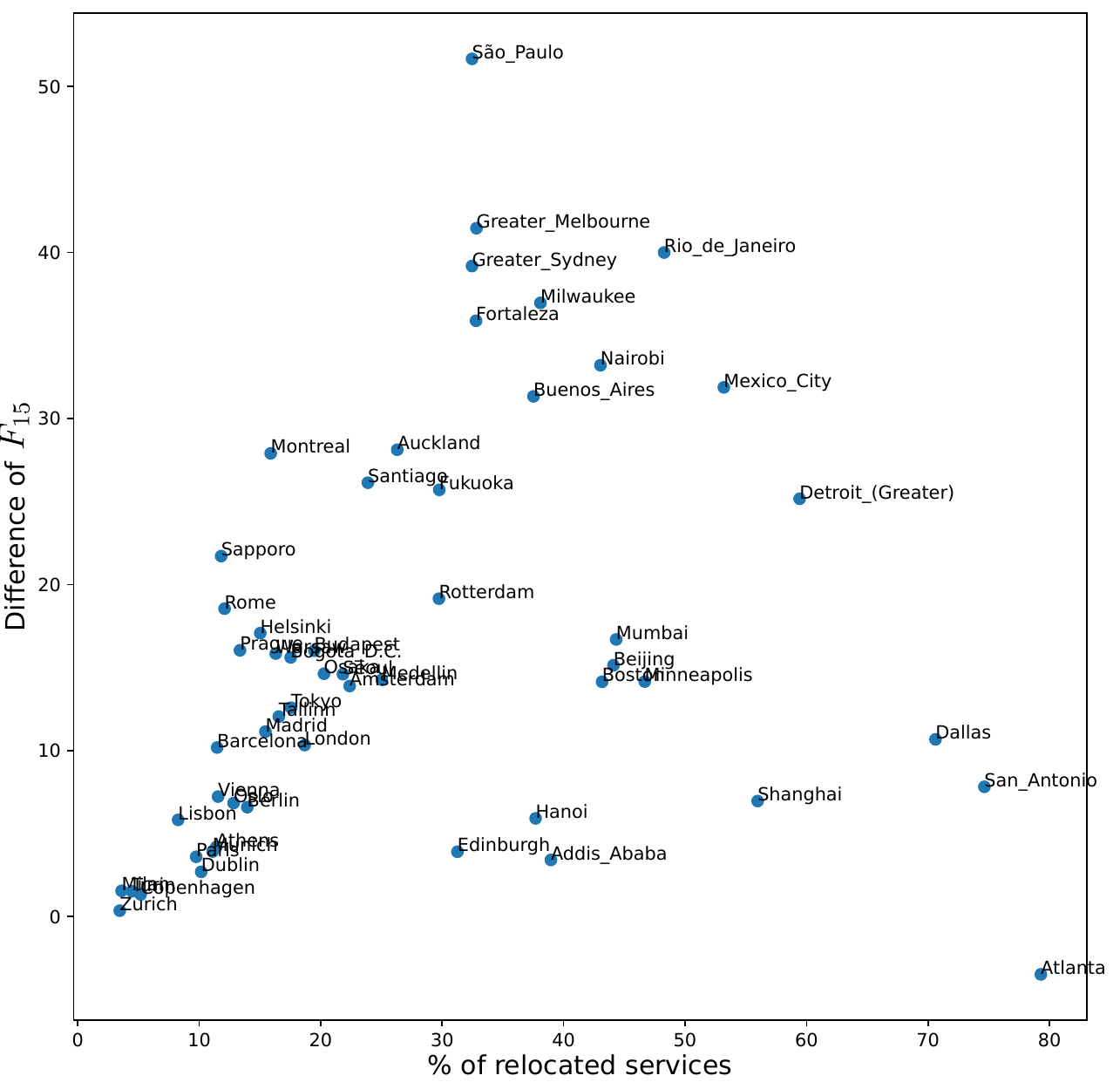}
    \caption{\textbf{Relocated percentage of services vs improvement in accessibility time.} The correlation between the percentage of services relocated by the algorithm and the accessibility improvements is present, but it is not straightforward. A higher correlation is present with the Proximity Time than with the percentage of 15-minute people.}
    \label{fig:si_reloc_vs_improvement}
\end{figure}

We show that the higher the percentage of services that are relocated, the higher the (positive) impact on the reduction of times of accessibility in cities: in Figure \ref{fig:si_reloc_vs_improvement} a scatter plot of the percentage of relocated services vs the accessibility improvement in minutes shows a positive correlation. All cities' average proximity times improve, and the percentage of people experiencing 15-minute accessibility increases in all cases except Atlanta. In Atlanta, the POIs are scarce with respect to its area and population. Thus, the relocation of the services makes the few 15-minute, most-served areas have fewer services, thus reducing the percentage of people with less than 15 minutes of proximity time.

\begin{figure*}[!htb]\label{fig:si_city_changes_rankings}
\includegraphics[width=.95\textwidth]{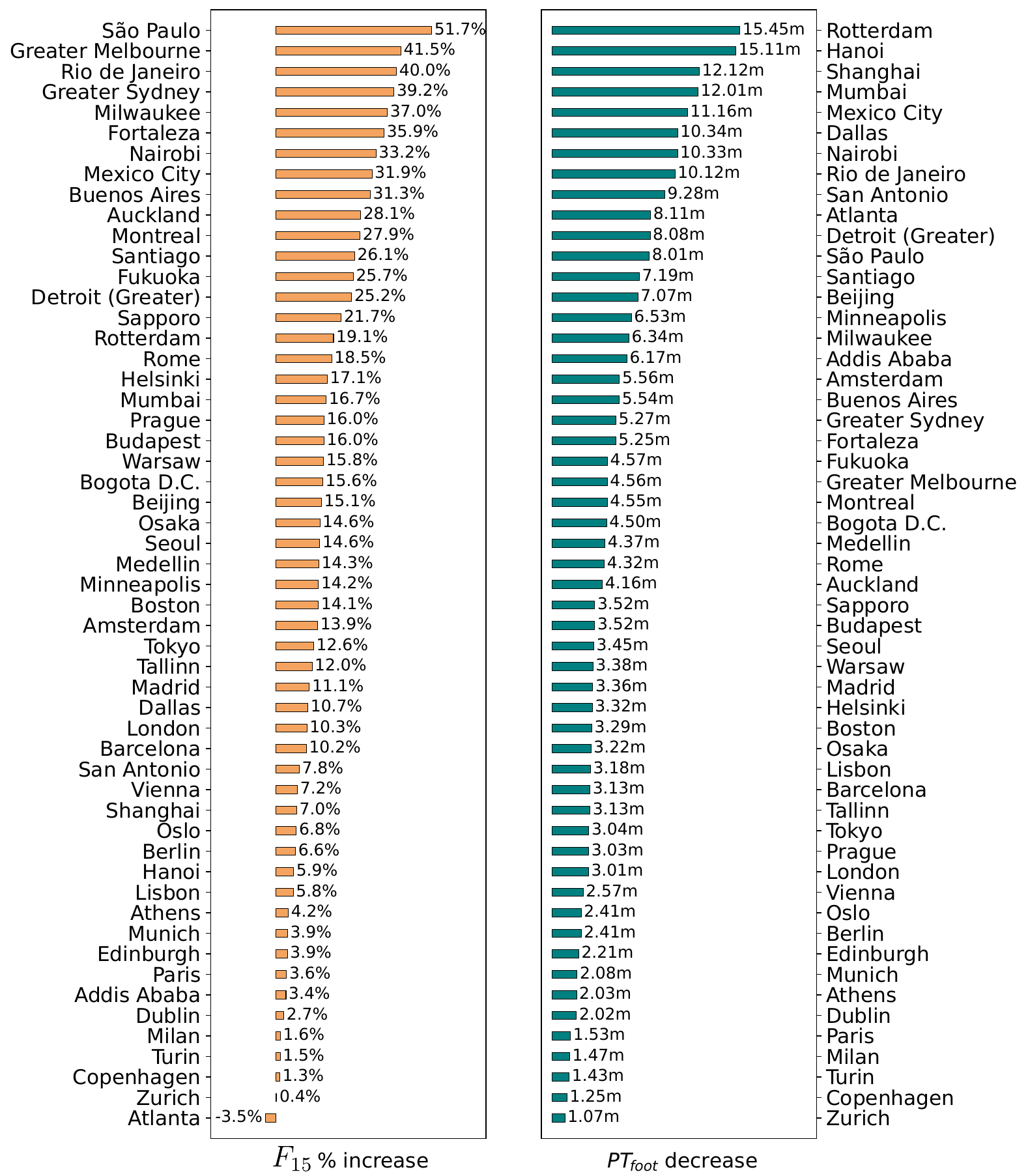}
\caption{\textbf{The improvements of the accessibility metrics in the optimal POI configuration.} On the left are the rankings of the percentage of people with 15-minute accessibility for the cities of our study. On the right is the improvement of the average Proximity Time of the city.} 
\end{figure*}

In Figure \ref{fig:si_city_changes_rankings}, we present the rankings of the changes in the accessibility scores of the set of cities we considered in our study. The changes are very heterogeneous, and the improvements in the percentage of people within 15-minute accessibility are not always reflected by the same change in the city's average Proximity Time.

\clearpage

\clearpage

\section{POIs vs proximity time scaling}

\begin{figure}[!htb]\label{fig:si_pois_scaling}
\centering
\includegraphics[width=.49\textwidth]{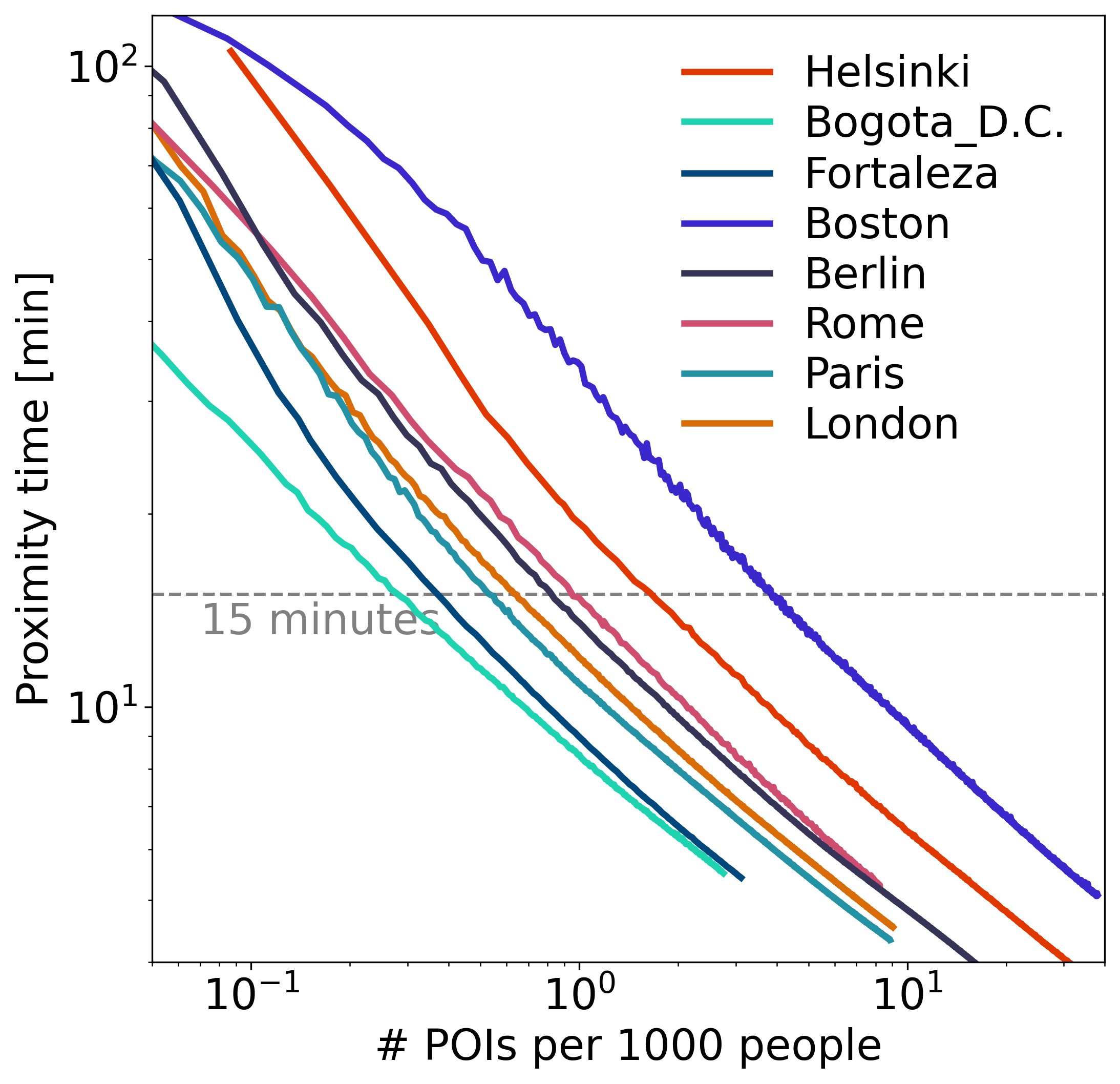}
\caption{\textbf{Number of optimally allocating POIs vs proximity time for selected cities.} Increasing the number of POIs optimally allocated, the proximity time decreases following a power law with an exponent of $\sim0.5$.} 
\end{figure}

The scaling of the Proximity Time average of the city with the number of POIs, when optimally placed, is shown in Figure \ref{fig:si_pois_scaling}. The figure is in a log-log scale and shows a power-law scaling with an exponent of 0.5. Again, the scaling of Boston's accessibility is separate from the other cities, like the analogous figure in the main text, meaning that its urban sprawl makes it difficult to realise the 15-minute city. 

\section{Optimally adding POIs to achieve the \textit{15-minute city}}

\begin{figure*}[!htb]\label{fig:si_opt_add_pois}
\includegraphics[width=\textwidth]{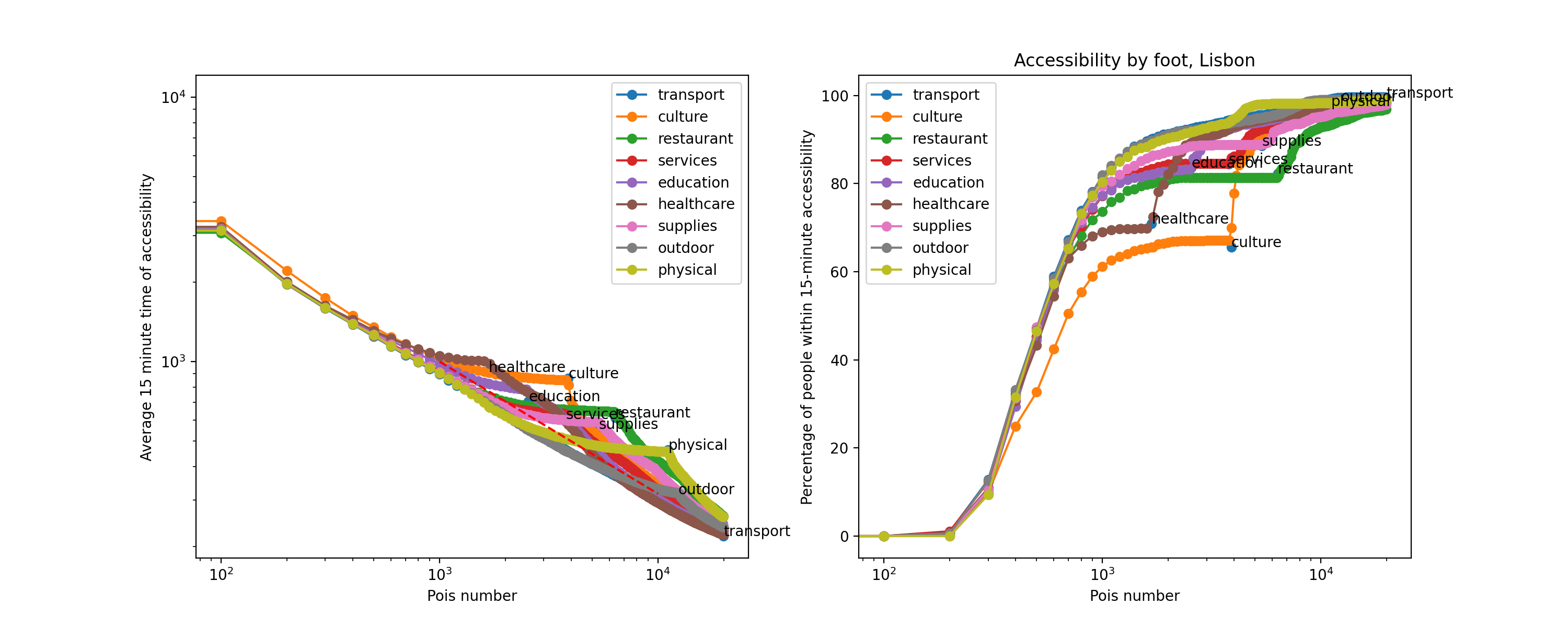}
\caption{\textbf{Adding or removing POIs from current scenarios in Lisbon.} The starting points, denoted by the text, are the real accessibility and POIs number of the category in the city.}
\end{figure*}

Our methodology can also be employed to understand how to remove or add POIs to the ones already present in a city. In Figure \ref{fig:si_opt_add_pois}, we present the results of this simulation for the city of Lisbon: it is interesting to notice that by adding POIs strategically, the percentage of people within 15-minute increases very fast while removing them where they are not needed the percentage stays roughly constant: for a policy-maker, this result implies that by fostering activities in a few selected areas the accessibility of the city will quickly improve greatly while moving away some activities from possibly overcrowded areas can have little impact on the overall city accessibility.

Additionally, it can be noted that different categories are far or close to the "optimal" line: the furthest seems to be the category of cultural activities, which are hard to distribute optimally, while the closest is the category of transport, which is placed well with respect with the resident population.

\section{How close to equality are cities?}

\begin{figure*}[!htb]
    \centering
    \includegraphics[width=.5\textwidth]{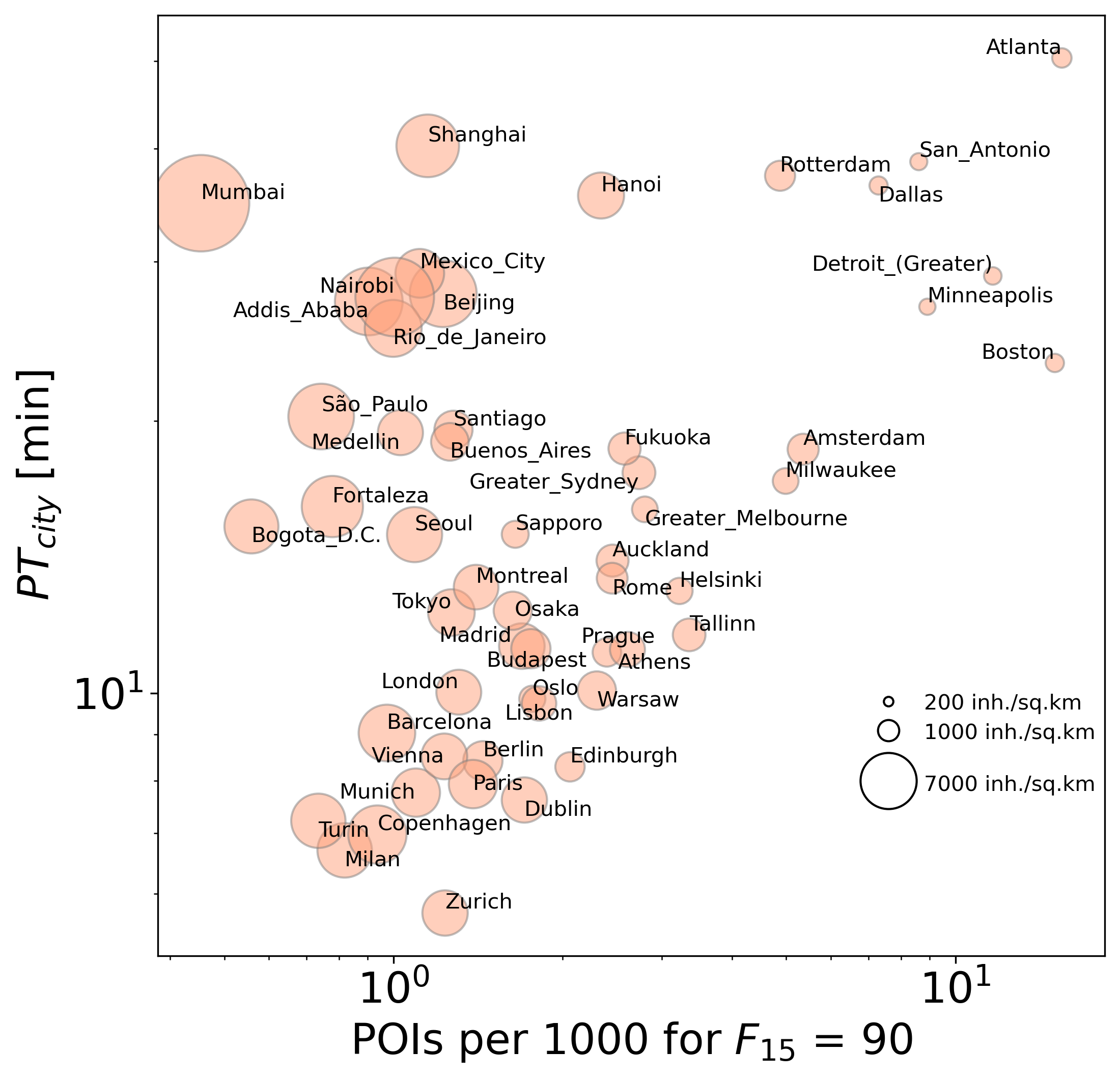}
    \caption{\textbf{Capacity required for 15-minute city vs current average accessibility time.}  The size of the points describes the population density of the cities.}
    \label{fig:si_cap_vs_avg}
\end{figure*}

The POIs needed for a 90\% 15-minute city describe how easy it is for a city to become 15-minute. This number can be compared with the actual accessibility score to understand which cities could reasonably improve their accessibility and which ones are already doing some of their best. In Figure \ref{fig:si_cap_vs_avg}, we show no clear trend: American cities are again outliers, needing many services and having bad accessibility. Some cities from less developed countries have the same accessibility but could easily improve by increasing the number of available services, although this metric can be biased by the incompleteness of POIs data.

\clearpage

\section{The number of POIs in cities}

\begin{figure}[!htb]
    \centering
    \includegraphics[width=.49\textwidth]{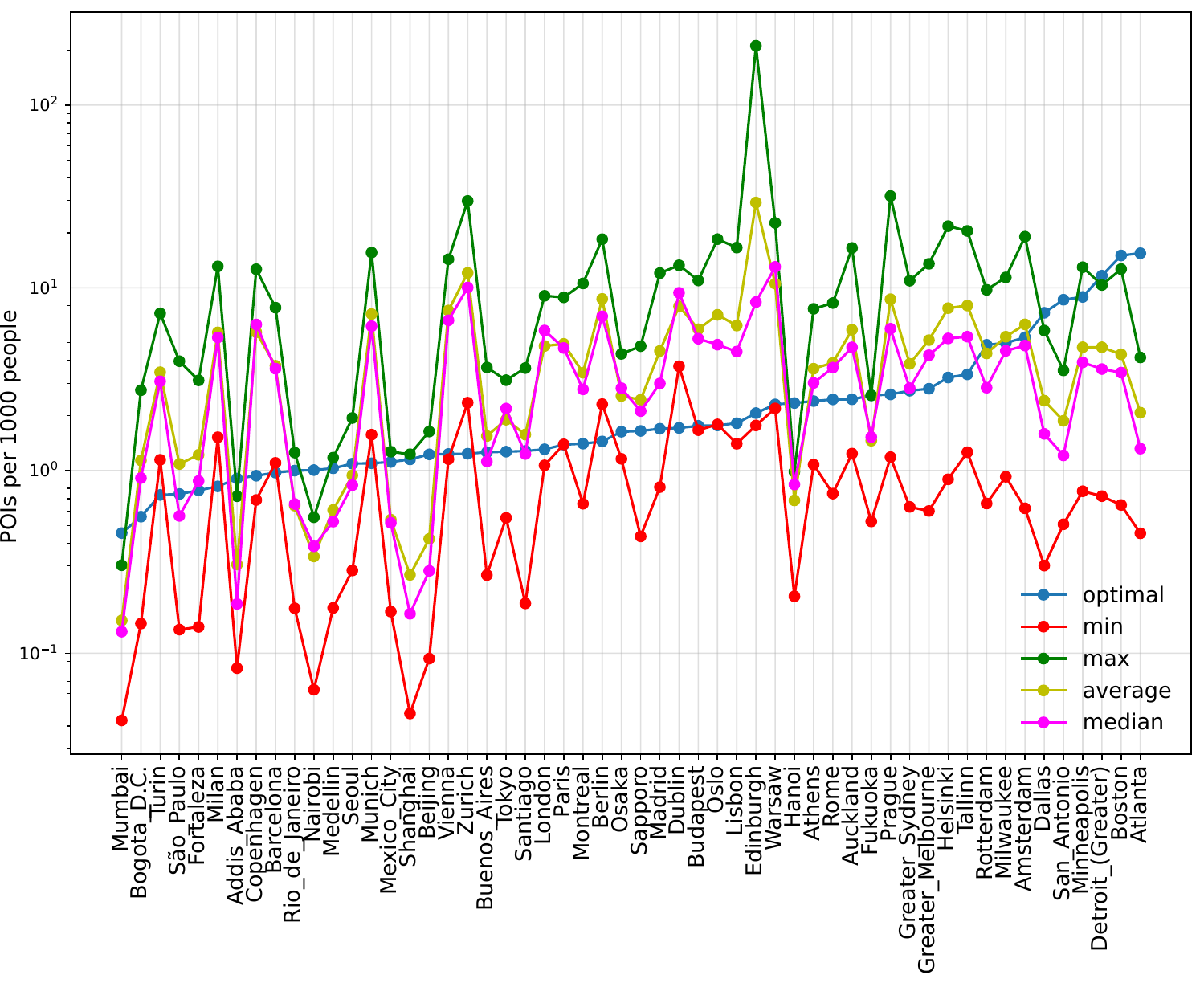}
    \includegraphics[width=.49\textwidth]{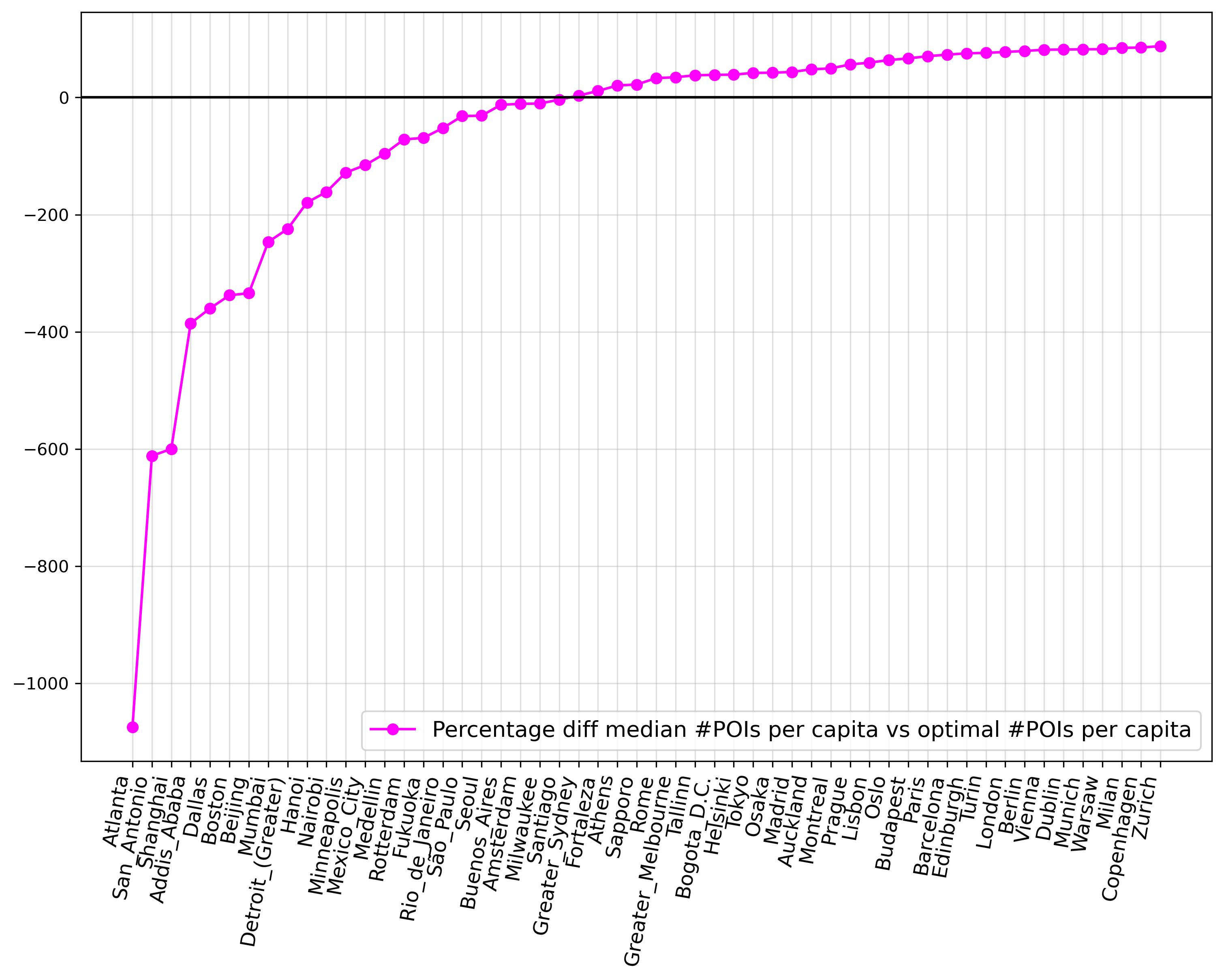}
    \caption{\textbf{Number of services vs optimal number of services (per 1000 people).} The top panel compares the optimal number of POIs per 1000 people in the city and the number of services present in the city, specifically the minimum amount of a category, the maximum, the average number of POIs per category and the median. The bottom plot shows the displacement in percentage of the median number of POIs for the categories vs the optimal number of POIs. Some cities have a very high number of services, while some have fewer services compared to what would be needed in the city.}
    \label{fig:si_num_pois}
\end{figure}

We can compare the optimal number of POIs per 1000 people to reach an inclusive 15-minute city with the number of POIs present in the city. For each category, the number is different, so in Figure \ref{fig:si_num_pois}, we can see the comparison between the minimum, maximum, average and median number of POIs of each category in the city with the one for the optimal 15-minute city obtained from our analysis. Interestingly, in many cases, this number lies between the minimum and maximum observed numbers, meaning that it is not unreasonable. The figure also describes the relative difference between the median observed and optimal numbers of POIs, constituting another metric of distance for cities to the ideal 15-minute city. Unsurprisingly, Atlanta is at the bottom also in this list, while Zurich sits atop.

% An appendix contains supplementary information that is not an essential part of the text itself but which may be helpful in providing a more comprehensive understanding of the research problem or it is information that is too cumbersome to be included in the body of the paper.

\end{appendices}

\end{document}